\documentclass[iop,twocolumn]{emulateapj}

\usepackage{epsfig, natbib} 
\usepackage{mathrsfs,amssymb}
\usepackage{graphicx,latexsym}

\newcommand{\etal}{{\rm et~al.\/}}

\newcommand{\kms}{\mbox{km\thinspace s$^{-1}$}}

\shorttitle{Dust-Poor B[e] Supergiants}
\shortauthors{Graus, Lamb, and Oey}

\begin{document}

\title{Discovery of New, Dust-Poor B[e] Supergiants in the Small
Magellanic Cloud}

\author{A. S.\ Graus,
         J. B.\ Lamb, and
	M. S.\ Oey}
\affil{Department of Astronomy, University of Michigan, 
830 Dennison Building, Ann Arbor, MI 48109-1042} 
\email{asgraus@umich.edu}
\slugcomment{Accepted 10 September 2012}

\begin{abstract}
We present the discovery of three new B[e] supergiants (sgB[e] stars) in the
Small Magellanic Cloud (SMC). All three stars (R15, R38, and R48) were
identified in the course of our Runaways and Isolated O Type Star
Spectroscopic Survey of the SMC (RIOTS4). The stars show optical
spectra that closely resemble those of previously known B[e] stars,
presenting numerous low-ionization forbidden and permitted emission
lines such as [Fe II] and Fe II.  Furthermore, our stars
have luminosities of $\log(L/$L$_{\odot})$ $\geq$ 4, 
demonstrating that they are supergiants.  However, we find lower
infrared excesses and weaker forbidden emission lines than for
previously identified B[e] supergiants.  Thus our stars appear to
either have less material in their circumstellar disks than other sgB[e]
stars, or the circumstellar material has lower dust content.
We suggest that these may constitute a new subclass of dust-poor sgB[e] stars.
\end{abstract}


\section{Introduction} \label{s:intro}

Massive stars comprise a small fraction of
the total number of stars in the universe, and are short-lived.
However, they provide a dominant portion of the mechanical and
radiative energy of galaxies.  Massive stars can enrich the universe
with metals, trigger star formation, and are a potential source
for the reionization of the universe.   The evolution of massive stars
is central to their ability to influence their surroundings.
Therefore, a better understanding of the phases of massive stellar
evolution will help us comprehend many physical processes that take
place within galaxies.  

In massive stars, the B[e] phenomenon is an evolutionary phase whose
origin remains unclear.  Stars exhibiting the B[e] phenomenon show
numerous emission lines, and normally show infrared radiation from  dust.
These properties point to the presence of large amounts of
circumstellar material, most likely in a disk configuration.  The B[e]
phenomenon is perplexing because stars in many different stages of
evolution can exhibit its signs.  \citet{b:Lamers98} divided the types
of stars showing the B[e] phenomenon into five groups: post-main
sequence B[e] supergiants (sgB[e] stars), pre-main sequence Herbig B[e]
stars, compact planetary nebulae, symbiotic B[e] stars, and an entire
category of unclassified B[e] stars.  A link between strong infrared
excess and B stars showing forbidden and permitted emission lines was
identified by \citet{b:Geisel70} and \citet{b:allenswings76}.  
\citet{b:Conti76} then suggested that ``B-type stars which show forbidden
emission lines'' should be classified as B[e] stars.  Since then,
many more B[e] stars have been found, leading to
a standard definition of the B[e] phenomenon as follows
\citep[e.g.,][]{b:Zickgraf98, b:Lamers98}:
strong Balmer emission lines; low-excitation permitted and
forbidden emission lines from low-ionization metals such as Fe; and
a strong excess in the near or mid-infrared due to circumstellar dust.
Among the more homogeneous and well studied group of stars
showing the B[e] phenomenon are the B[e] supergiants. 

Supergiant B[e] stars (sgB[e]) also have well defined criteria laid out by
\citet{b:Lamers98}, including supergiant luminosity
($\log L/L_{\odot}\geq$ 4), indicators of evolution off the main sequence
sequence, broad ultraviolet (UV) absorption lines, and signs 
of nitrogen abundance enhancement.  In order to explain the emission properties of sgB[e] stars,
\citet{b:Zickgraf85} proposed a two-component wind model where the star is rotating near break-up velocity and simultaneously
 close to the Eddington limit.  This causes the star to emit a
wind that is fast and sparse near the poles, producing the high excitation absorption, while being denser around
the equator, producing the emission features.  The two-component wind allows for the creation of a
dusty circumstellar disk, which produces the infrared excess seen
around sgB[e] stars.  In this scenario, the star begins exhibiting the
B[e] phenomenon sometime after the red supergiant phase, but before
He-core burning begins \citep{b:Lamers06}.  Some sgB[e] stars exhibit properties akin to those
of Luminous Blue Variables (LBVs) suggesting that LBVs and at least some
sgB[e] stars are related, although their relation remains unclear, as
summarized by \citet[]{b:Lamers06}.  R4 in the Small 
Magellanic Cloud (SMC) is a prominent candidate for this model.  R4 is
currently classified as a LBV/sgB[e] because it shows properties of
both objects.  \citet{b:Pasquali00} further suggested that R4 could
have formed in a binary merger.  It is unknown whether this is a common
origin for sgB[e] stars in general, or whether R4 is a unique case.
Others have suggested that sgB[e] stars arise from interacting
binaries, which is consistent with discoveries of binary companions
around sgB[e] stars
\citep[e.g.,][]{b:Wheelwright12, b:Miro03}.  Binary companions
have been found around other types of B[e] stars
as well, such as around Herbig B[e] stars
\citep[e.g.,]{b:Kraus11}.  This scenario could help explain why
stars exhibiting the B[e] phenomenon show similar properties while
apparently being at different stages of evolution. 

In this paper, we report the discovery of three previously
unidentified sgB[e] stars in the SMC:  R15, R38, and R48
\citep[]{b:Feast60}, along with observations of a 
previously known sgB[e], AzV 154 \citep[]{b:Azzopardi75}, obtained from the Runaways and
Isolated O Star Spectroscopic Survey of the SMC
\citep[RIOTS4;][]{b:Oey11, b:Lamb11}. 
The RIOTS4 survey is a photometrically selected and spatially complete spectroscopic survey
of OB stars in the SMC.  The three previously unknown sgB[e] stars all
show optical spectra characteristic of the B[e] phenomenon, including
forbidden and permitted emission.  These stars also show additional
indicators that they are supergiants, and they appear to be post-main
sequence stars.


\section{RIOTS4 Data } \label{s:MandO}

The RIOTS4 survey was carried out for a photometrically selected
sample of field OB stars in the SMC \citep{b:Oey11, b:Lamb11}.  OB
stars were selected from a $UBVR$ photometric survey of the SMC
carried out by \citet{b:Massey02}.  The most massive O stars, along with 
B stars of roughly type B1.5 and earlier, were selected from these
data by requiring that $B$ $\leq$ 15.21 and the reddening free parameter
$Q_{UBR}\leq -0.84$.  \cite{b:Oey04} then identified field stars in this
sample with the use of a friends-of-friends algorithm
\citep[][]{b:Battinelli91}.   Stars were considered clustered if
another massive star was found within a given clustering length of
the target star.   For this survey the clustering length was 28 pc.
The field stars selected by the friends-of-friends algorithm were then
observed using the Inamori Magellan Areal Camera and Spectrograph
(IMACS) and Magellan Inamori Kyocera Echelle (MIKE) spectrograph on
the Magellan telescopes at Las Campanas Observatory. 

Over the course of the survey we discovered four stars showing rich emission-line
spectra:  R38, R48, R15, and AzV 154.  Of these stars,
AzV154 is a known sgB[e] \citep{b:Zickgraf89}.  R48 was observed
using the IMACS multi-slit mode on the Magellan Baade
telescope, using the f/4 camera with a 1200 lines/mm grating
and a 0.7$\arcsec$ slit, yielding a spectral resolution of $R\sim$ 3600, and a wavelength range
of 3650 \AA~to 5250 \AA.  
R38,  R15, and AzV 154 were observed with the MIKE 
echelle spectrograph on the Magellan Clay telescope.  These
observations were made using a 1$\arcsec$ slit, yielding a resolution of $R\sim 28,000$, and a wavelength
range of 3250 \AA~to 5050 
\AA. 

For the IMACS data, the reduction, bias subtraction, wavelength calibration, and flat
fielding of the 2-d spectra were performed using the COSMOS\footnotemark[1] data reduction pipeline.  The
spectrum was extracted using the $apextract$ package in IRAF\footnotemark[2].  For data obtained using
the MIKE spectrograph, the data were reduced and wavelength-calibrated
using IRAF packages.  The 1-d spectra from
both instruments were then rectified and cleaned of cosmic rays and bad pixels using
IRAF.

\footnotetext[1]{http://code.obs.carnegiescience.edu/cosmos}

\footnotetext[2]{IRAF is distributed by the National Optical Astronomy Observatories,
    which are operated by the Association of Universities for Research
    in Astronomy, Inc., under cooperative agreement with the National
    Science Foundation.}

\bigskip\bigskip
\section{Optical Properties} \label{s:optical}

\subsection{Optical Spectra} \label{s:spectra}

\begin{figure*}
\includegraphics*[height=108mm, width=170mm]{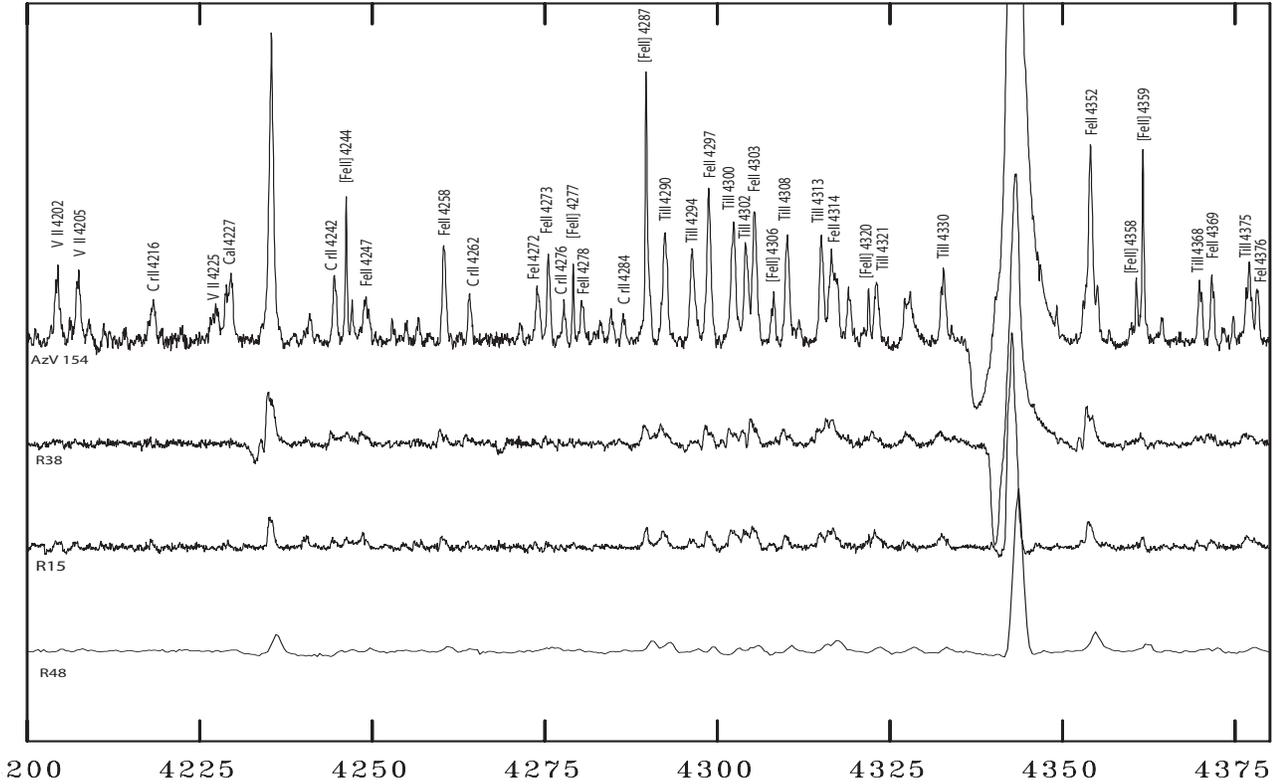}
\caption[]{Optical spectra of the region near Fe II $\lambda$4233
  and H$\gamma$, for the previously known  
  sgB[e] star AzV 154, along with those of the newly discovered sgB[e]
  stars R38, R15, and R48 (top to bottom).  The top three spectra were
  obtained with the MIKE echelle spectrograph, and the bottom spectrum
  with IMACS.  } 
\label{f:spectra}
\end{figure*}

Optical spectra of our three, newly-identified sgB[e] stars, along
with that of the previously known object, 
are shown in Figure~\ref{f:spectra}, and their properties are given in
Table~\ref{t:optical}.  Table~\ref{irphotometry} gives their photometry.
Our optical spectra clearly show strong emission features, 
particularly in the Balmer series.  R38 exhibits strong P Cygni
profiles in its Balmer lines as does the previously known sgB[e] star, AzV~154,
indicating intense mass loss.  In accordance with the standard
definition of the B[e] phenomenon presented in, e.g.,
\citet{b:Zickgraf98} and \citet{b:Lamers98}, our stars also show numerous other
emission lines.  These lines are primarily permitted and forbidden
low-ionization lines from metals, such as Fe II, [Fe II], Ti II, 
Cr II, and V II.  We identified emission lines in the spectral range 4000 --
4750 \AA, based on the work of \citet{b:Zickgraf89, b:AJ5, b:AJ4,
  b:AJ3} and \citet{b:AJ1}.  Four to twelve
forbidden emission lines are confirmed in each of the three new stars
at these wavelengths, including [Fe II] $\lambda\lambda$4288, 4360, 4414, and [Ni II]
$\lambda$4462, which are detected in all objects.  We list the heliocentric
radial velocities (RV) in Table~\ref{t:optical}, together
with the root-mean-square velocity residual $\langle v_{\rm rms}\rangle$ measured
for the fitted RV from both permitted and forbidden lines, as well as that
measured from only the forbidden lines, $\langle v_{\rm fb,rms}\rangle$.  Given that
there are fewer forbidden lines, the values of $\langle v_{\rm
  fb,rms}\rangle$ are fully consistent those of $\langle v_{\rm rms}\rangle$.

The absorption lines of B-star spectra are also present, for example,
He I, Si II, and Mg II.  The classification criteria of
\citet{b:Fitzpatrick91}, and \citet{b:Walborn90} were used 
for spectral typing of our stars (Table~\ref{t:optical}).
The presence in absorption of Si II $\lambda\lambda$4128-30, and the
relative strength of Mg II $\lambda$4481 compared to He I $\lambda$4471
indicate the spectral types of 
all three of our newly discovered sgB[e] stars to be in the mid to
late B range.  These spectral types are unexpectedly cooler
than those derived for normal OB stars in the RIOTS4 survey.
While it is possible that a cooler binary companion dominates the
optical spectra, it turns out that higher $R$-band fluxes caused by the
strong H$\alpha$ emission in the decretion disks actually lead to lower $Q_{UBR}$ values,
mimicking a bluer continuum.  This reveals a selection bias for the
RIOTS4 survey which thus includes other $R$-excess objects, in addition
to the target blue stars, thereby also selecting a large number of
classical Be stars.

\begin{deluxetable*}{l c c c c c c c c c c}
 \tablewidth{0pt}
 \tabletypesize{\tiny}
 \tablecaption{RIOTS4 SMC sgB[e] Stars \label{t:optical}}
 \tablehead{\colhead{\citet{b:Massey02}} &
	\colhead{AzV\tablenotemark{a}} &
	\colhead{Radcliffe\tablenotemark{b}} &
	\colhead{RA (J2000)} &
	\colhead{Dec} &
	\colhead{SpT\tablenotemark{c}} &
	\colhead{$\log(L/L_{\odot}$)} &
	\colhead{$v\sin i$\tablenotemark{d}} &
	\colhead{RV\tablenotemark{d}} & 
	\colhead{$\langle v_{\rm rms}\rangle$\tablenotemark{d}} &
	\colhead{$\langle v_{\rm fb,rms}\rangle$\tablenotemark{d}}}
\startdata
29267 & AzV 154 & \nodata & 00:54:09.52 & -72:41:43.3 & B0 I & $5.23\pm0.15$ & \nodata & 
   141.0 & 3.1 & 3.4 \\
46398 & AzV 230 & R15 & 00:59:29.20 & -72:01:04.6 & B7 I & $4.79\pm0.17$ & 56 & 
   126.4 & 9.9 & 16.6 \\
62661 & AzV 390 & R38 & 01:05:47.01 & -71:46:21.9 & B8 I & $4.61\pm0.15$ & 53 & 
   105.0 & 7.3 & 15.2 \\
83480 & \nodata & R48 & 01:30:10.89 & -73:18:56.2 & B6 I & $4.36\pm0.12$ & 108 & 
   202.5 & 10.0 & 13.9 
\enddata
\tablenotetext{a}{Identification from \citet{b:Azzopardi75}.}
\tablenotetext{b}{Identification from \citet{b:Feast60}.}
\tablenotetext{c}{Spectral type for AzV 154 is
   from \citet{b:Zickgraf89}; those for R15, R48, and R38 are new.}
\tablenotetext{d}{\kms.}
\end{deluxetable*}

We can use the photospheric absorption lines to estimate the
projected rotational velocity of our stars (Table~\ref{t:optical}) by
gaussian fitting of the Si II, Mg II, and He I lines.  The values given
in the Table are rough estimates that correspond directly to the
measured FWHM, since the accurate measurement of $v\sin i$ is difficult;
effects such as
macroturbulence can cause significant, additional line broadening
for supergiants \citep[]{b:Diaz07}.  Of
course another important factor is the inclination angle with which we
are viewing our stars.  Two of our stars (R15 and R38) show
single-peaked Balmer emission lines.  If the Balmer emission comes
from a disk, single peaked lines imply that we are viewing these two objects
close to pole-on \citep[]{b:Stee94}, so their rotational velocities
could be significantly higher.  This therefore suggests that
macroturbulence significantly affects our measurements of the
projected rotational velocity in Table~\ref{t:optical}.

We calculated the stellar luminosities from their $V$-band fluxes,
using photometry from \citet{b:Massey02} (Table~\ref{irphotometry}),
along with bolometric corrections for B supergiants from
\citet{b:Cox00}.  These luminosities may be upper limits since there
may be some contribution from the circumstellar disk. 
Photometry from the All Sky Automated Survey
\citep[ASAS;][]{b:Pojmanski97} hints that our three new, sgB[e] stars may
be variable at a level of $\sim0.1$ mag in $V$, and the known sgB[e]
star AzV 154 shows variability in $V$ at a level of 0.25 mag.
\citet{b:Lamers98} notes that sgB[e] generally have low photometric
variability.  Extinction values were estimated using the
 extinction maps of \citet{b:Zaritsky02}, and a distance modulus of 
18.9 was used \citep[]{b:Harries03}.  These quantities are also given
in Table~\ref{irphotometry}.  

\begin{figure}[!t]
\includegraphics[height = 80mm, width = 80mm]{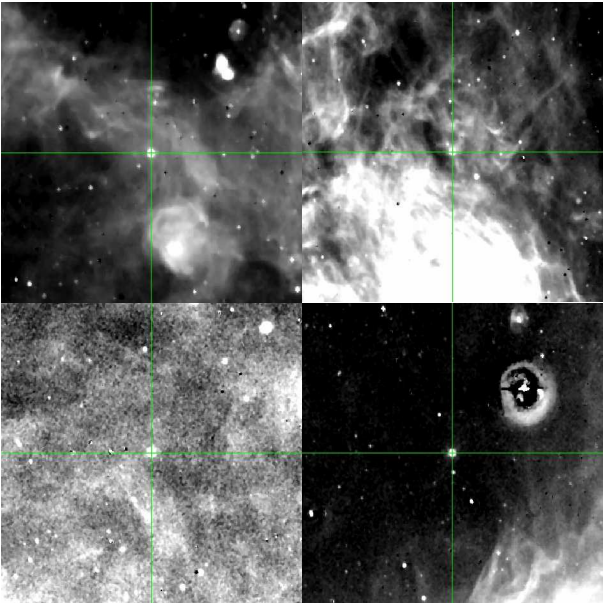}
\caption[]{MCELS H$\alpha$ images of the regions around
  our stars (clockwise, from top left): 
  AzV 154, R15, R38, and R48. The top left image is 116 pc across, the
  rest are 186.6 pc across.} 
\label{f:Halpha}
\end{figure}

\begin{figure}[!t]
\includegraphics[trim = 5mm 3mm 0 0, height = 75mm, width = 85mm]{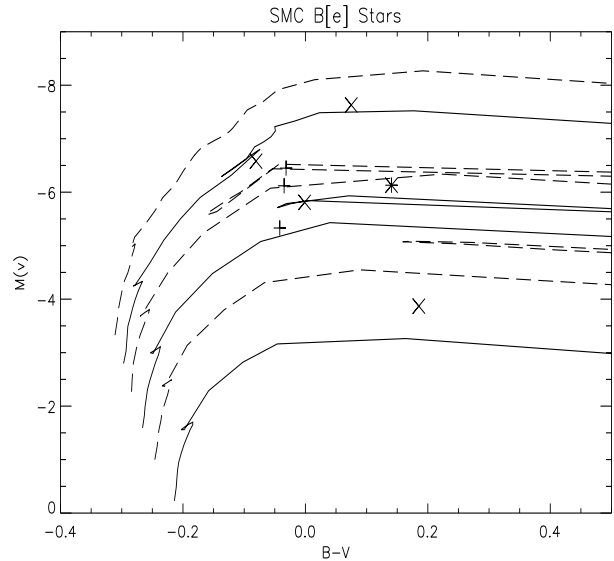}
\caption[]{A color-magnitude diagram showing R15, R48, and R38 (plus
  signs), AzV 154 (asterisk), and the other SMC sgB[e] stars from the
  literature \citep{b:Zickgraf06, b:Wisniewski07} (Xs).  The mass
  tracks were calcuated for SMC metallicites of stars
  \citep[]{b:Charbonnel93} with masses of 20, 12, and 7M$_{\odot}$
  (dashed lines), and 15, 9, and 5M$_{\odot}$ (solid lines).} 
\label{f:hrdiagram}
\end{figure}

\begin{deluxetable*}{l c c c c c c c c c c c c c}
  \tablewidth{0pt}
  \tabletypesize{\tiny}
  \tablecaption{Photometry\tablenotemark{a} \label{irphotometry}}
  \tablehead{\colhead{Star} &
	\colhead{$A_{V}$\tablenotemark{c}} &
	\colhead{$U$} & 
	\colhead{$B$} & 
	\colhead{$V$} & 
	\colhead{$R$} &
	  \colhead{$J$} &
	  \colhead{$H$} &
	  \colhead{$K$} &
	  \colhead{[3.6]} &
	  \colhead{[4.5]} &
	  \colhead{[5.8]} &
	  \colhead{[8.0]} &
	  \colhead{[24]}
	}
\startdata
AzV 154 & $0.24\pm0.21$ & 12.80 & 13.48 & 13.20 & 12.64  & 12.35 & 11.93 & 11.11 & 9.18 & 8.34 & 7.75 & 6.97 & 4.79 \\
R15 & $0.17\pm0.21$ & 11.81 & 12.67 & 12.64 & 12.46 & 12.37 & 12.28 & 12.10 & 11.51 & 11.29 & 10.10 & 10.63 & 9.30 \\
R38 & $0.16\pm0.18$ & 12.24 & 13.01 & 12.98 & 12.82 & 12.68 & 12.59 & 12.38 & 11.97 & 11.77 & 11.41 & 11.17 & \nodata \\
R48 & $0.34\pm0.26$ & 12.92 & 13.74 & 13.73 & 13.54 & 13.45 & 13.34 & 13.18 & 12.79 & 12.57 & 12.38 & 11.94 & \nodata \\
VFTS 698\tablenotemark{b} & \nodata & \nodata & 14.12 & 13.68 & \nodata & 12.34 & 11.89 & 11.43 & 10.60 & 10.23 & 10.02 & 9.66 & \nodata 
\enddata
\tablenotetext{a}{$UBVR$ photometry is taken from \citet{b:Massey02},
  $JHK$ from 2MASS \citep{b:Skrutskie06}, and 3.6 -- 24 $\mu$m from the
  SAGE-SMC survey \citep{b:Gordon10}.}
\tablenotetext{b}{Photometry for VFTS 698 from \citet{b:Dunstall12}.}
\tablenotetext{c}{Visual extinction values taken from \citet{b:Zaritsky02}} 
\end{deluxetable*}

\subsection{Supergiant B[e] or Herbig B[e]?}

 We now use the data we have compiled to constrain the evolutionary state of our
three newly discovered B[e] stars.  The selection criteria of the
RIOTS4 survey only identify
luminous, early-type stars.  This means that we can immediately rule out
the B[e] phenomenon due to planetary 
nebula sources or symbiotic stars.  Thus, the stars must either be
pre-main sequence objects or evolved supergiants.
At most, about 25\% of massive stars in the SMC form outside clusters
\citep[]{b:Oey04}, and therefore the RIOTS4 field star criteria select
against the pre-main sequence Herbig B[e] stars.
To further verify that our stars are not present in star
forming regions, we searched for dense nebulae by checking images
of our stars in H$\alpha$, [SII], and [OIII] using data from the
Magellanic Clouds Emission-Line Survey \citep[MCELS;][]{b:Smith98, b:Pellegrini12}
Figure~\ref{f:Halpha} shows that although there is 
nebular emission present near R15, none of the stars are inside regions
clearly suggestive of current, active star formation.

We then compare our stars to the criteria of sgB[e] stars laid out by \citet{b:Lamers98}. 
Our stars are all supergiants, having luminosities above the limit
suggested by Lamers of log($L$/L$_{\odot}$) $\geq$ 4
(Table~\ref{t:optical}).  AzV 154 and R38 both show P Cygni
profiles in the Balmer recombination lines, which is a secondary
criterion of sgB[e] stars.  

We directly compare the optical spectra of our three newly-identified sgB[e] stars
to that of the previously discovered sgB[e] star, AzV 154.  While the emission lines of
AzV 154 have a much higher intensity than for our stars, they show many of
the same lines, particularly the Fe II and [Fe II] lines typical of sgB[e] stars
(Figure~\ref{f:spectra}).  
Figure~\ref{f:hrdiagram} shows a color-magnitude diagram (CMD)
for our three stars (plus signs), along with AzV 154 (asterisk).
The crosses show the other known SMC sgB[e] stars from the literature.
Figure~\ref{f:hrdiagram} indicates  
that our stars are clearly off the main sequence, as shown 
by the evolutionary tracks of \citet{b:Charbonnel93}.
The proximity of these other SMC sgB[e] stars to our stars
suggests that they are near the same evolutionary phase. 
Thus, our stars are clearly evolved supergiants.

\begin{figure}[!t]
\includegraphics[height= 80mm, width = 80mm]{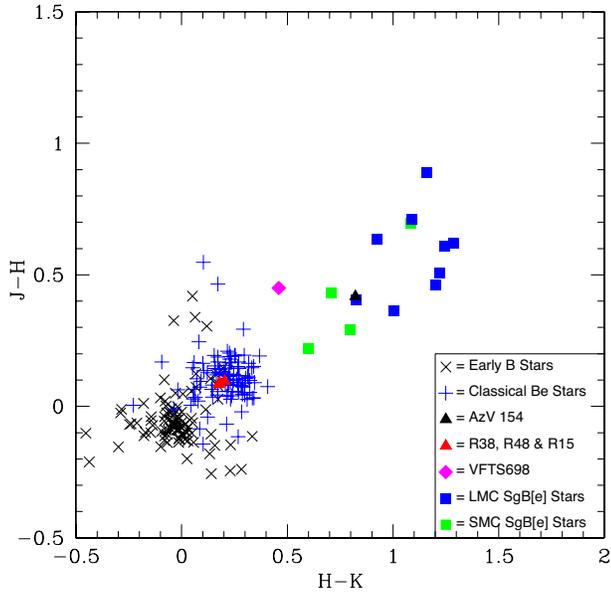}
\caption[]{Near-IR color-color diagram ($J-H$ vs $H-K$) comparing R15,
  R48, and R38 with SMC B and Be stars, and previously known sgB[e]
  stars in the Magellanic Clouds.  This figure
  demonstrates that our new objects show IR excess typical of classical Be
  stars and atypical of most other sgB[e] stars in both the LMC and
  the SMC.} 
\label{f:nearir}
\end{figure}

\begin{figure}[!t]
\includegraphics[height = 80mm, width = 80mm]{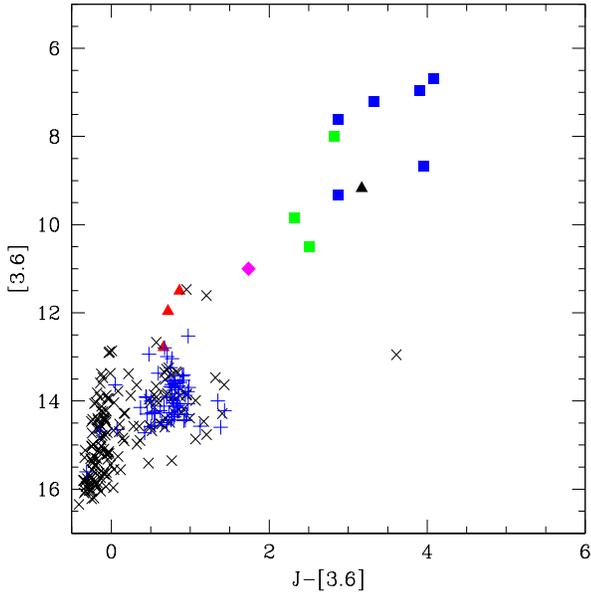}
\caption[]{Color-magnitude diagram of [3.6] vs $J-$[3.6], again
  showing the disparity between our stars and known sgB[e] stars.
  The [3.6] magnitudes for the LMC stars are adjusted to the SMC
  distance by the 0.4 mag difference in their distance moduli.
  Symbols are the same as in Figure \ref{f:nearir}.} 
\label{f:midir}
\end{figure}

\bigskip\bigskip
\section{A Lack of Dust Emission} \label{s:IR}

\subsection{Infrared Data}

An important property of B[e] stars is a large infrared
excess due to circumstellar dust.  SgB[e] stars are noted for their
combination of high luminosity and strong infrared excess.  An example
of this is \citet{b:Bonanos10} who used data from the Spitzer SAGE SMC
survey to study the properties of massive stars in the SMC.  In all
infrared color-magnitude diagrams and color-color diagrams, the sgB[e]
stars form a clearly distinct group of massive stars.  We obtained
$JHK$ photometry of all the B-type stars, including the Be and sgB[e]
stars, in the RIOTS4 survey from the Two Micron All Sky Survey
\citep[2MASS;][]{b:Skrutskie06}, and 3.6 -- 24 $\mu$m photometry from
the SAGE SMC survey \citep[]{b:Gordon10}.  The IR photometry for our
objects is given in Table~\ref{irphotometry} and plotted in
Figures~\ref{f:nearir} and \ref{f:midir} relative to these other
groups of SMC B stars.  The Figures also show the previously known
sgB[e] stars in the SMC and LMC \citep{b:Zickgraf06,b:Wisniewski07},
which show a large IR excess, whereas our stars are on average 0.6 mag
bluer than AzV 154 in $H$-$K$, 0.3 mag bluer in $J$-$K$, and 2.4 mag
bluer in $J$-[3.6].  The modest IR excess for our stars, R15, R38, and R48,
is similar to that for classical Be stars.  In contrast, B stars that do not show 
evidence of a circumstellar disk have IR colors around zero, with the exception
of some outliers which are just evolving off the main sequence.

\begin{figure}[!t]
\includegraphics[height=80mm, width=85mm]{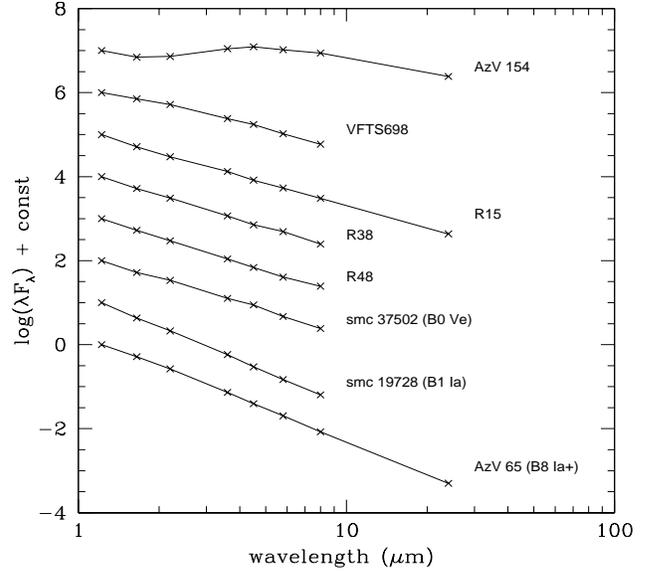}
\caption[]{IR SEDs of R15, R38, R48, and AzV154, together with that
  for VFTS 698, a dust-poor sgB[e] star in the LMC.  The SEDs are normalized to
  their $J$-band magnitudes.  For comparison, we also 
  show a classical B0 Ve star (smc 37502), a  B1a supergiant (smc 19728), and a late B8Ia+ supergiant
\citep[AzV 65;][]{b:Bonanos10}}
\label{f:irsed}
\end{figure}

Figure~\ref{f:irsed} shows the infrared spectral energy
distributions (SEDs) for the sgB[e] stars, using the same data sets.
For comparison, we also show the SEDs for three other stars from the RIOTS4 survey
including a representative classical Be star, smc 37502 \citep{b:Massey02}; and OB
supergiant, smc 19728, along with a late-type B supergiant (AzV 65)
taken from \citet{b:Bonanos10}. 
Again, our three stars are very clearly different from the previously known
sgB[e] star, AzV 154.  However, they are also
different from the B supergiants, whose SED's are steeper.  Interestingly, the
infrared properties of our three 
stars closely match the infrared properties of classical Be stars.
This is apparent in all three of Figures~\ref{f:nearir}, \ref{f:midir},
and \ref{f:irsed}.

\subsection{Low Infrared Excess}

Through the infrared data obtained from the 2MASS and SAGE SMC surveys, we
see that our stars show an infrared excess that is distinctly lower than that of previously
discovered sgB[e] stars.  This suggests a lack of warm dust in the
circumstellar material around our three stars.
An excess in $K-$[24] $\geq$ 5 indicates the presence
of an optically thick, dusty disk \citep[]{b:Hernandez06}.  Our stars fall well below
that limit, around $K-$[24] = 2.8, indicating that a dusty
circumstellar disk is lacking, but indicating the presence of an
optically thin circumstellar disk, which is characterized by $K-$[24] $\geq$ 0.3. 

The modest infrared excess of our stars is similar to
that of classical Be stars (Figures \ref{f:nearir} and \ref{f:midir}), and
their IR SEDs are also similar (Figure~\ref{f:irsed}). 
Figure~\ref{f:irsed} shows that R15, R38, and R48 lack the IR bump from hot dust
emission typical of sgB[e] stars like AzV~154.   
While, being supergiants, our stars are much brighter, the general shape of the
SEDs of our stars more closely follows that of classical Be stars.  
Thus, the IR emission in our objects may not be
due to dust emission as is typical of sgB[e] stars, but may instead result from free-free
emission as in classical Be stars \citep[]{b:Gehrz74}.  In the IR,
free-free opacity increases at longer wavelengths, leading to an
inverse power-law SED \citep[]{b:Lamers99}. 
For example, \citet{b:Kastner10} obtained IR spectra of several sgB[e] stars and
found evidence of free-free emission in the near IR.  Thus
our stars appear to have optically thin circumstellar disks, since
they show weak IR excesses.  

The lack of circumstellar dust could be due to a couple different
factors.  First, the circumstellar disks around our stars could be
ultra metal-poor.  The SMC is already known to be a metal-poor environment,
and the normal sgB[e] stars in the SMC show a lower infrared excess on 
average than those in either the LMC or the Galaxy.  Alternatively,
our sgB[e] stars simply could have much smaller, or less massive,
circumstellar disks, possibly because
they are just transitioning into or out of the sgB[e] phase.  If there
is a lack of material in the disks around our stars, then the disks would
be optically thin, meaning that dust would not be able to form, and
the only infrared excess would be due to free-free emission.  
If our stars have such low-mass circumstellar disks, this would manifest
in the optical spectra through the presence of weak emission lines, as
opposed to the very strong emission lines seen in other sgB[e] stars;
this is exactly what is seen in our spectra. 

\subsection{A new class of sgB[e] stars?}

We have referred to R15, R38, and R48, as sgB[e] stars
in this work.  However, as described earlier, the current
definition of B[e] stars includes a ``strong'' infrared excess (e.g.,
Conti 1997; Zickgraf 1998).  This has been straightforward, since the
optical spectral properties and infrared emission have always appeared
to occur together.  Now, however, our discovery of several objects having
optical B[e] spectra, while lacking the strong IR excess, presents a
complication for classification.  While \citet{b:Lamers98} specifies the
criteria for B[e] stars to include the strong IR excess, he also
states that the spectroscopic criteria were defined for the  optical
spectrum.  On the advice of Peter Conti and Henny Lamers, we classify
these IR-weak objects as B[e] stars, based on the original, optically-based 
definition of the stars (P. Conti; H. J. G. L. M. Lamers, private
communications).  Our classification therefore reverts to
a definition of B[e] stars based exclusively on the existence of B star
properties together with optical permitted and forbidden emission lines
\citep{b:Conti76}.  This is reinforced by the fact that the B[e]
nomenclature references only the optical spectral features.

Our three objects have optical and infrared properties that are
quantitatively more similar to each other than to other sgB[e] stars.
And, this is not the first detection of objects showing all of the
properties of sgB[e] stars, but 
lacking the characteristic dusty infrared excess.  Recently,
\citet{b:Dunstall12} reported the detection of a similar object in 30
Doradus in the Large Magellanic Cloud (LMC).  This object, VFTS 698,
shows optical [Fe II] emission and IR properties similar to the three
stars that we have found, and to classical Be stars.  As also found in
our stars, the photospheric emission is visible in VFTS 698, as
seen in the the absorption-line spectrum of He I and Si II; whereas the
photosphere is only occasionaly
seen in sgB[e] stars.  We include the data for this star in  
Table~\ref{irphotometry} and Figures~\ref{f:nearir} -- \ref{f:irsed}.
There are likely to be a number of other such known stars, for
example, MWC 314 in the Milky Way \citep[]{b:Miroshnichenko96,
  b:Miroshnichenko98}.  We caution that both VFTS 698 and MWC 314 are
known binaries, which may complicate the comparison with our newly
found SMC objects.  However, one of our objects, R48, shows an anomalously high
radial velocity (Table~\ref{t:optical}) which may be evidence of binarity.
Thus, dust-poor sgB[e] stars are found in widely different environments:  VFTS698 in 30
Doradus, our three stars in the field of the SMC, and MWC 314 in the
Milky Way.  We therefore propose that such stars may be a distinct
subset of sgB[e] stars, which may be transitioning to or from
the full-fledged B[e] phenomenon.

\section{Conclusions} \label{s:conc}

We report spectroscopic observations of three new sgB[e] stars, along
with similar observations of AzV 154, a previously known sgB[e] star.  These
stars were observed in the course of the RIOTS4 survey of massive field
stars in the SMC.  Our stars show rich emission-line spectra,
including emission in the Balmer lines, and forbidden and permitted emission
from low-excitation metals typical of the B[e] phenomenon.  The
stars have supergiant luminosities.  Based on observations of our stars 
in H$\alpha$, [SII], and [OIII], they are apparently not associated
with star-forming regions, implying that they are unlikely to be
Herbig B[e] stars.  The presence of P Cygni profiles in one of the spectra,
the comparison to AzV 154, and the position of our stars on the H-R
diagram are also consistent with the properties of sgB[e] stars. 

Most interestingly, our stars do not show the strong IR excess
characteristic of other sgB[e] stars.  Instead, the IR
emission from our stars resembles that of classical Be stars,
suggesting that it originates from free-free emission.  Thus, our 
stars lack the strong circumstellar dust disks that produce
the IR excess in B[e] stars.  This could be due to low dust content in
the disks, or it could be due to the disks being much lower mass.
The existence of other dust-poor sgB[e] stars
\citep[e.g.,][]{b:Dunstall12, b:Miroshnichenko98}
suggests that such objects are a distinct sub-category of sgB[e] stars, which
may be transitioning toward or away from the normal sgB[e] phase.

\acknowledgments

We thank Nuria Calvet, Peter Conti, Joel Kastner, Henny Lamers, and
Selma de Mink for useful discussions.  We are also grateful to the
anonymous referee and to Alceste Bonanos for comments on the manuscript. 
Thanks to Eric Pellegrini for providing nebular data and to
Jordan Zastrow for editorial help.  Funding for this project was
provided by NSF grant AST-0907758.   


\clearpage

\end{document}